\documentclass[aps,prl,reprint,superscriptaddress]{revtex4-1}
\usepackage{blindtext}
\usepackage{amsopn,amsmath,amssymb,amsfonts}
\usepackage{mathtools}
\usepackage{graphicx,subfigure}
\usepackage{xcolor}
\usepackage[normalem]{ulem}

\graphicspath{{.}{./Results/}}

\begin{document}

\title{On the problem of non-zero word error rates for fixed-rate error correction codes in continuous variable quantum key distribution}
\author{Sarah J. Johnson}
\email{sarah.johnson@newcastle.edu.au}
\affiliation{School of Electrical Engineering and Computer Science, The University of Newcastle, Australia}
\author{Andrew M. Lance}
\affiliation{QuintessenceLabs Pty. Ltd., Canberra ACT,  Australia.}
\author{Lawrence Ong}
\affiliation{School of Electrical Engineering and Computer Science, The University of Newcastle, Australia}
\author{Mahyar Shirvanimoghaddam}
\affiliation{School of Electrical Engineering and Computer Science, The University of Newcastle, Australia}
\author{T. C. Ralph}
\affiliation{Centre for Quantum Computation and Communication Technology, School of Mathematics and Physics, University of Queensland, Brisbane, Queensland 4072, Australia}
\author{Thomas Symul}
\affiliation{QuintessenceLabs Pty. Ltd., Canberra ACT,  Australia.}

\begin{abstract}

The maximum operational range of continuous variable
quantum key distribution protocols has shown to be
improved by employing high-efficiency forward error
correction codes. Typically, the secret key rate model
for such protocols is modified to account for
the non-zero word error rate of such codes.
In this paper, we demonstrate that this model is incorrect:
Firstly, we show by example that fixed-rate error correction
codes, as currently defined, can exhibit efficiencies greater than unity. Secondly,
we show that using this secret key model combined with
greater than unity efficiency codes, implies that it is possible to
achieve a positive secret key over an entanglement breaking channel - an impossible scenario.
We then consider the secret key model from a post-selection perspective, and examine the implications for key rate if we constrain the forward error correction codes to operate at low word error rates.

\end{abstract}

\maketitle

\section{Introduction}

Quantum key distribution is one of the most advanced applications of quantum physics and information science. It enables the distribution of information-theoretically secure random key material between two parties in spatially-separated locations connected by an unsecured optical link \cite{Gisin_2002}.

There are two complementary approaches to quantum key distribution: discrete variable quantum key distribution uses single-photon or weak coherent states and single photon detectors \cite{Scarani_2009_RevModPhys_QKD}, while continuous variable quantum key distribution (CVQKD) uses coherent or squeezed states of light and homodyne detectors \cite{Weedbrook_2012}. Both discrete and continuous quantum key distribution systems have been demonstrated (for a review see \cite{Lo_2014}) and importantly both the discrete and continuous approaches to quantum key distribution have been proven to be information-theoretic secure, the latter against collective attacks \cite{Garcia-Patron_2006,Navascues_2006} and with composable security \cite{Leverrier_2015}.

Continuous variable quantum key distribution has gained interest recently because of the potential  technology advantages it offers that may enable higher secret key rates. Technological advantages include high-quantum efficiency homodyne detectors; high-speed commercial-off-the-shelf optical components and compatibility with optical network infrastructure.

Originally limited to short distances \cite{Grosshans_PRL_2002}, the operation range of CVQKD protocols was considerably extended by employing the reverse reconciliation protocol that exploits one-way communication, including forward error correction codes in the error reconciliation post processing step \cite{Grangier03}. Low density party check (LDPC) and multi-edge LDPC (ME-LDPC) codes are examples of high-efficiency forward error correction codes that have been employed in CVQKD systems \cite{Lodewyck2007_LDPC_CVQKD,Fossier_CVQKD_NJP2009,Leverrier_2009,Jouguet-2011-multiedge,Jouguet2012_experimental,Huang_CVQKD2016}. The ME-LDPC codes in particular exhibit good error correction performances at low signal-to-noise (SNR) ratios, making them suitable for CVQKD applications.

A simple and useful model for the secret key rate of CVQKD protocols in the case of
collective attacks can derived when one assumes a specific reconciliation procedure.
Specifying the reconciliation procedure to be forward error reconciliation,
the secret key rate can then be empirically modeled by \cite{Leverrier_phd2009}
\begin{equation} \label{eq:Leverrier_key_rate_2}
\Delta I = \beta I_{\rm AB} - I_{\rm E}
\vspace{-0.5em}
\end{equation}
where $I_{\rm E}$ denotes the bound on
an eavesdropper's (Eve's) maximum accessible information,
$I_{\rm AB}$ denotes the capacity of the channel between
the sender (Alice) and the receiver (Bob), and $\beta$
denotes the efficiency of reconciliation.

The model for the secret key rate in Eq.~(\ref{eq:Leverrier_key_rate_2})
is commonly used in the literature to compare the performance of CVQKD
protocols under varying conditions, for example, different amounts of loss and noise.
Importantly, Eq.~(\ref{eq:Leverrier_key_rate_2}) assumes
that that every codeword of the forward error correction code is decoded correctly.

Practical applications of ME-LDPC and LDPC codes inherently exhibit a non-zero word error rate (WER)
\cite{Lodewyck2007_LDPC_CVQKD,Martinez-Mateo_WER_DVQKD_2013}.
The WER is the rate at which the decoder fails to
decode the correct codewords. Since the adoption of forward error correction codes
in CVQKD systems, the model of the secret key rate
has been subsequently modified to include the efficiency and WER of the code \cite{Lodewyck2007_LDPC_CVQKD,Huang_CVQKD2016}
\vspace{-0.5em}
\begin{equation} \label{eq:key_rate}
\Delta I = (\beta I_{\rm AB} - I_{\rm E})(1-p_\mathrm{fail})
\vspace{-0.5em}
\end{equation}
where $p_\mathrm{fail}$ is the rate at which the decoder fails to
decode to a valid codeword. We point out that it is not clear if this
secret key model was formally derived.  We also note that for all error correction schemes there is always a non-zero probability that a valid, but incorrect, codeword could be returned. I.e. the decoder will fail but will not know that it has failed. This would not be detected by Alice and Bob, resulting in an incorrect secret key which would not be discovered until the key is used. In our simulations, we have used the true word error rate (as we have knowledge of the transmitted message so can detect all failures). In practice, this won't be possible, however it it usually assumed that the undetected error rate is negligible.

In this paper we show that the current key model
Eq.~\eqref{eq:key_rate} is incorrect.
Firstly, we demonstrate that fixed-rate error correction codes
can exhibit efficiencies, $\beta$ greater than unity for a range of word
error rates. Secondly, we show that by using the secret key model
Eq.~\eqref{eq:key_rate} combined with $\beta$ greater than unity,
it is possible to achieve a positive secret key over
an entanglement breaking channel. We then consider the secret key model from a CVQKD post-selection perspective, and also examine the implications for key rate if we constrain the forward error correction codes to operate at low word error rates.

\section{forward error correction codes
with $\beta$ greater than unity} \label{sec:Background}

Since the adoption of forward error correction
codes in CVQKD protocols, the model for
secret key rate has been formulated to account for the non-ideal
performance of the forward error correction coding scheme.
The two sources of this non-ideal performance are
1) the losses due to mapping a binary error correction code
onto a Gaussian-input channel, and
2) the losses due to the performance of a practical error correction
code compared to the ideal, capacity achieving, error correction code.

It has been shown \cite{Grosshans_2003} that entanglement-based
CVQKD protocols are equivalent to so-called
prepare-and-measure CVQKD protocols where Alice transmits
Gaussian modulated coherent states. In the latter case, if the sender
encodes with Gaussian signals, a mapping protocol can be used to transform
the Gaussian symbols to binary symbols for subsequent error correction.
Recent advances have been made on efficient mapping protocols
(for example see \cite{Leverrier_2008} and references therein),
which have been shown to be highly efficient  for
low SNR  ratios, typically required for
CVQKD  protocols \cite{Leverrier_2008}.
We denote the efficiency of the mapping protocol, compared to perfect Gaussian
mapping by $\beta_\mathrm{MAP}$. This mapping efficiency,
although important, is not the focus of this paper.

Following the mapping step, reconciliation is then performed
using an error correction code designed for the binary-input additive Gaussian white noise (BI-AWGN)  channel,
in particular LDPC \cite{Lodewyck2007_LDPC_CVQKD,Fossier_CVQKD_NJP2009}
and ME-LDPC codes \cite{Leverrier_2009,Jouguet-2011-multiedge,Jouguet2012_experimental,Huang_CVQKD2016}.

The efficiency of the forward error correction code is calculated according to \cite{Lodewyck2007_LDPC_CVQKD,Jouguet-2011-multiedge,Leverrier_phd2009}
\begin{equation} \label{eqn:efficiency}
\beta_\mathrm{FEC} = \frac{R}{I_{\rm AB}}
\end{equation}
where $R=k/n$ is the rate of the error correction
code that maps a $k$ bit message to an $n$ bit codeword,
and $I_{\rm AB}$ denotes the capacity of the channel between
the sender (Alice) and the receiver (Bob), i.e. the rate of the
ideal capacity-achieving code.
The total efficiency of the reconciliation scheme is then
\begin{equation} \label{eqn:total_efficiency}
\beta =  \beta_\mathrm{MAP} \beta_\mathrm{FEC}.
\end{equation}

In the remainder of this paper we will assume
that $\beta_\mathrm{MAP}=1$. Substituting Eq.~\eqref{eqn:efficiency}
into Eq.~\eqref{eq:key_rate} we see that the key
rate for a rate $R=k/n$ forward error correction code operating
at a  WER of $p_\mathrm{fail}$ can be calculated as
\begin{equation} \label{eq:key_rate_short}
\Delta I = (R - I_{\rm E})(1-p_\mathrm{fail}).
\vspace{-0.5em}
\end{equation}

For LDPC and ME-LDPC codes, the code performance at a given
SNR (where the SNR is determined by the parameters of the
CVQKD protocol model) can be
determined theoretically by evaluating the expected performance of an
ensemble of infinite length codes with a particular structure using density
evolution \cite{Richardson_irregLDPC}. Density evolution returns
the \textit{threshold} of a rate $R$ code ensemble, which is the smallest
SNR at which the BER, the fraction of codeword bits
remaining in error following decoding, goes to zero as decoding proceeds.
For a given SNR, we find the highest code rate $R$ with a threshold
at or below that SNR. Then $\beta$ is \cite{Jouguet-2011-multiedge}
\begin{equation} \label{eqn:efficiency_th}
\beta_\mathrm{FEC} = \frac{R}{I_{\rm AB}(s_\mathrm{th})}
\end{equation}
where $I_{\rm AB}(s_\mathrm{th})$ is the rate of
the capacity achieving code on an AWGN channel with SNR equal to $s_\mathrm{th}$.
The secret key rate model with a rate $R$ code with  threshold $s_\mathrm{th}$ is
\begin{equation}
\Delta I =  R - I_{\rm E}(s_\mathrm{th}),
\end{equation}
where $I_{\rm E}(s_\mathrm{th})$ is
the bound on the information leaked to Eve with SNR equal to $s_\mathrm{th}$ and the WER is zero. By the definition of the threshold, the code operates at a zero error rate and consequently
$R \leq I_{\rm AB}(s_\mathrm{th})$ follows from the channel
coding theory and so $\beta_\mathrm{FEC} \leq 1$.

In contrast, for any finite length code, the error rate is
bounded away from zero. Instead, the performance of finite-length
error correction codes can be determined via simulation to find
the error rate curves (BER and WER) for a particular code as
the SNR is varied. Since the error rate decreases as the SNR
is increased, the SNR value at which a given code will be
operated, $s_\mathrm{op}$, will depend on the WER that can be tolerated by the application.
Thus the rate $R$ of the error correction code we can employ
on a given channel will depend on the WER we allow.

For example, the rate $0.02$ code in \cite{Jouguet-2011-multiedge} has a theoretical threshold
(found using density evolution) of $s_\mathrm{th} = 0.02865$.
Consequently,
\begin{equation}
\beta_\mathrm{FEC} = \frac{R}{I_{\rm AB}(s_\mathrm{th} = 0.02865) } = \frac{0.02}{0.02038} = 0.981,
\end{equation}
and the  secret key rate model
with SNR $s = 0.02865$ is
\begin{equation}
\Delta I = 0.02 - I_{\rm E}(s=0.02865).
\end{equation}
In practice, a length $2^{20}$ rate $0.02$ code operating
at an SNR of $s=0.029$ has a WER of $1/3$ \cite{Jouguet-2011-multiedge}.
Applying Eq.~\eqref{eqn:efficiency}, this gives
\begin{equation}
\beta_\mathrm{FEC} = \frac{R}{I_{\rm AB}(s_\mathrm{op}= 0.029)} = \frac{0.02}{0.02062} = 0.97,
\end{equation}
 The secret key rate  model for this code operating
 at a WER of $1/3$ with SNR $s=0.029$ is
\begin{equation}
\Delta I = 2/3\Big(0.02 - I_{\rm E}(s=0.029)\Big).
\end{equation}

\begin{figure}[h!]
    \centering
    \includegraphics[width=\columnwidth]{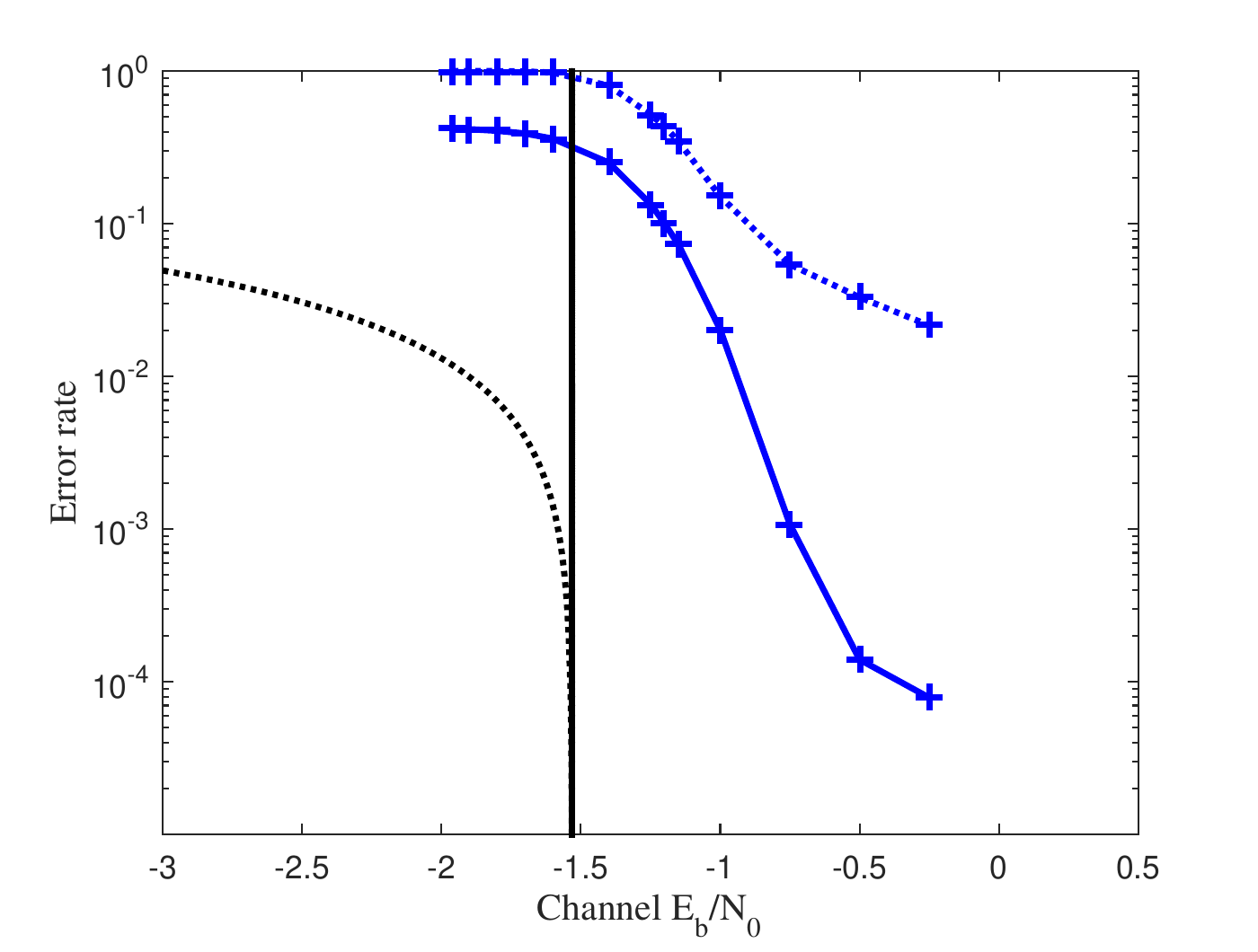}
\includegraphics[width=\columnwidth]{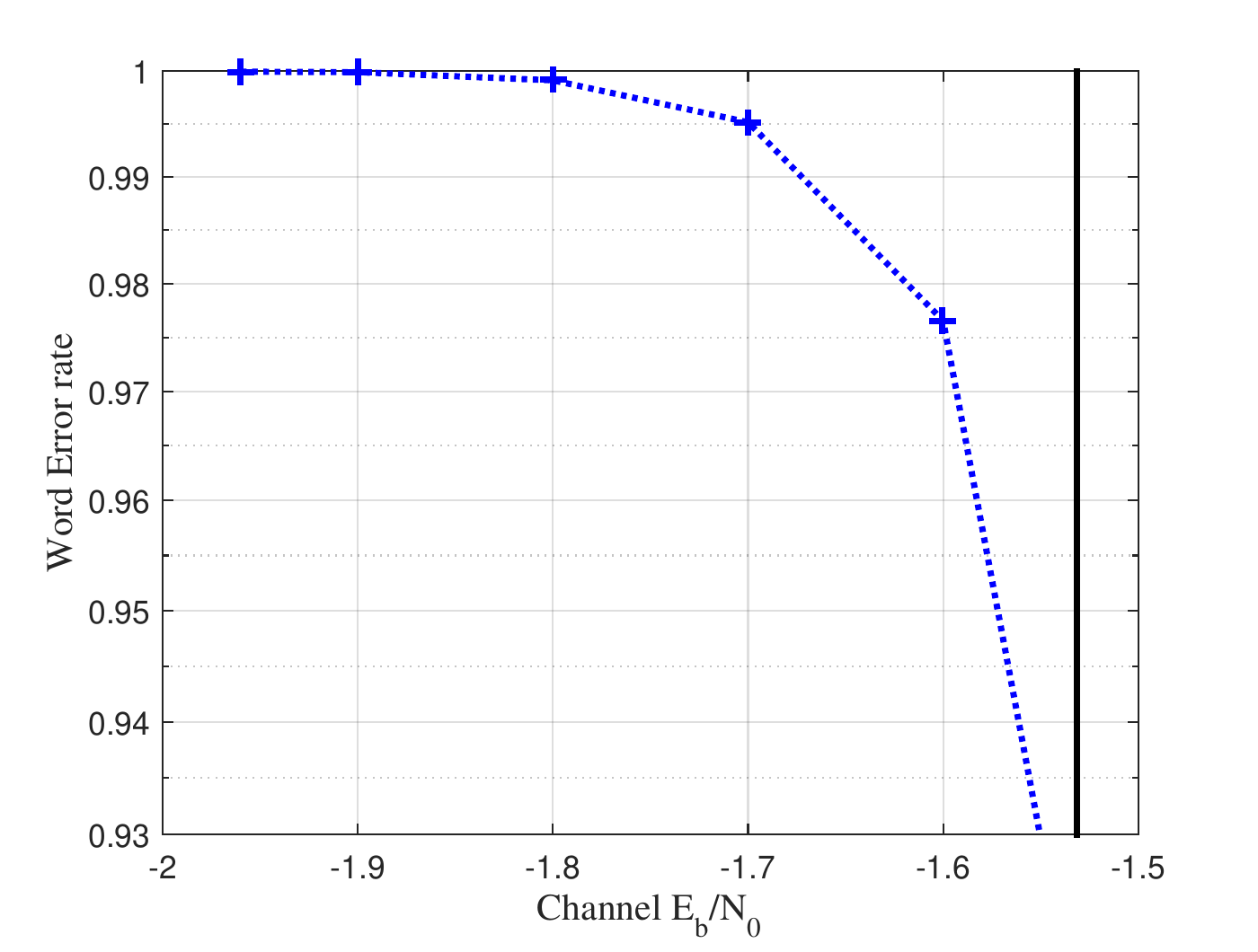}
    \caption{  BER (blue solid curve)
    and  WER (blue dashed curve) for a
    length $10^5$ rate $0.02$, ME-LDPC code with degree
    distribution from \cite{Jouguet-2011-multiedge} with
    a maximum of $5,000$ decoder iterations.
    Here $R$ is the code rate and $s$ is the channel SNR.
    Also shown is the zero-error channel capacity for this
    code rate (solid black curve) and the non-zero error
    channel capacity for this code rate (dashed black curve).
    \label{fig:ME-code-rate-1-50-results-v2c}}
\end{figure}

If we take the practice of allowing non-zero WERs to the extreme, it is possible to significantly increase
the rate $R$ we can operate at a given SNR, $s$ to above $I_{\rm AB}(s)$,
or equivalently, to operate at a $\beta$ above $1$.
For example, Fig.~\ref{fig:ME-code-rate-1-50-results-v2c} shows
the finite length performance of a rate $0.02$ code with
degree-distribution from \cite{Jouguet-2011-multiedge} when the code
length is $10^5$ bits and a maximum of $5,000$ decoder iterations are allowed.
 By allowing a WER of $0.9999$, one can operate with rate $0.02$
at an SNR of $0.0258$ and thus
\begin{equation}
\beta_\mathrm{FEC} = \frac{R=0.02}{I_{\rm AB}(s_\mathrm{op}= 0.0258)} = \frac{0.02}{0.018374} = 1.09.
\end{equation}
 The secret key rate model for this code operating
 at a WER of $0.9999$ with SNR $s=0.0258$ is
\begin{equation}
\Delta I = 0.00001 \Big(0.02 - I_{\rm E}(0.0258)\Big).
\end{equation}
For the same code rate, $R$, increasing the WER has allowed
us to reduce $s$ and thus reduce $I_{\rm E}(s)$. Consequently,
the current secret key
rate model informs us that by increasing the WER we are able to
increase the range of $s$ for which the  same rate $R$ code can operate.

\begin{figure}[h!]
    \centering
    \includegraphics[width=\columnwidth]{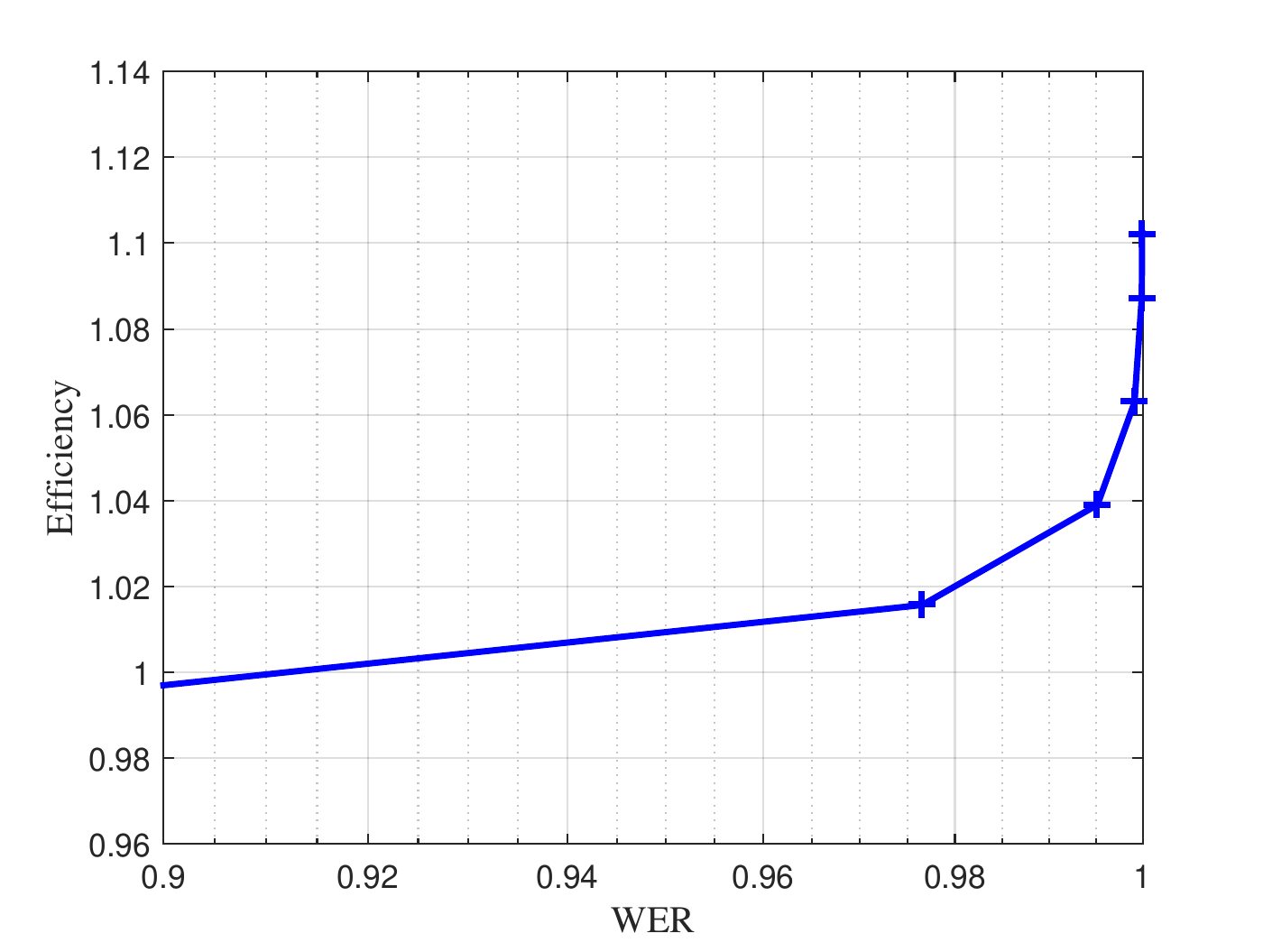}
    \caption{ The values of $\beta$, Eq.~\eqref{eqn:efficiency}, obtained for the
    length $10^5$ rate $0.02$, multi-edge LDPC code with degree
    distribution from \cite{Jouguet-2011-multiedge} with a
    maximum of $5,000$ decoder iterations.
    \label{fig:ME-code-rate-1-50-efficiency}}
\end{figure}

Also shown in Fig.~\ref{fig:ME-code-rate-1-50-results-v2c}
is the Shannon capacity result for rate $0.02$ codes and
we can see that the ME-LDPC code is in fact operating at an
SNR below the Shannon channel capacity which is why $\beta$ can be calculated as greater than $1$. Of course this code is not outperforming the Shannon channel capacity,
rather the comparison is not valid. Coding schemes with a
non-zero error rate are not bound by the same capacity formula
as schemes with a zero error rate. It is well known that the capacity
of the AWGN channel varies as the BER is allowed to
increase \cite{Aldis_capacity92}. If the  BER is below
$10^{-4}$ the effect is insignificant \cite{Aldis_capacity92},
however, above this, the SNR required to decode at the given BER reduces
significantly if a non-zero error rate is allowed.
Also shown in Fig.~\ref{fig:ME-code-rate-1-50-results-v2c} is the
non-zero error-rate capacity. Indeed, if we consider that we are
allowed to operate at error rates as high as we like, the results
in \cite{Aldis_capacity92} tell us that by operating close to the
non-zero-error-rate capacity we can obtain a very large increase in SNR over the zero-error-rate capacity and thus obtain a $\beta >> 1$.
Fig.~\ref{fig:ME-code-rate-1-50-efficiency} shows
$\beta$ results we have obtained in practice for the code
from Fig.~\ref{fig:ME-code-rate-1-50-results-v2c} as the
WER is varied.

In this section we have demonstrated that fixed-rate
forward error correction codes operated
at a non-zero WER can operate at lower SNR than ideal codes
operating at a zero WER, thus returning a $\beta$ above unity.
Alternatively, if the SNR is fixed, operating the codes
at higher WERs increases the code rate $R$ that can be used thereby
increasing the likelihood that we can obtain a positive $(R - I_{\rm E})$
term in the secret key model. In the following section we
demonstrate that this has important consequences for the validity of
the key rate model, as currently defined.\\
\\

\section{Codes with $\beta$ greater than unity applied to
unity-gain classical teleporter quantum channels}

We have demonstrated that is it possible to operate
at a code rate above that of an ideal code by using,
fixed-rate forward error correction codes operating at sufficiently
high WERs. We will now
describe a problem that arises with the secret
key model in Eq.~\eqref{eq:key_rate} when considering
codes operating at a non-zero WER.

In quantum information theory, it has been shown that when a
quantum channel is replaced with a classical teleporter
\cite{furusawa1998unconditional},
no entanglement can be transmitted
through such a channel \cite{Grosshans_2003}.

As a consequence, in
the context of quantum key distribution, no secret key can be established
through such a quantum channel \cite{Lutkenhaus_2004}. Such a channel
would be equivalent to an intercept-and-resend attack.

A Gaussian quantum channel
is completely described by the channel transmission and the
channel excess noise \cite{Grangier03}.

In the case where a quantum channel is replaced by a
classical teleporter operating at unitary gain \cite{furusawa1998unconditional},
the quantum channel parameters are described by a transmission
$T=1$ and a (relative input) excess noise of $\varepsilon=2$.

From Eq.~(\eqref{eq:key_rate}), we can see that the secret key model is positive when
\begin{equation}
\beta >\frac{I_{\rm E}}{I_{\rm AB}},
\end{equation}
this is equivalent to $R > I_{\rm E}$ from the previous section.
We can therefore calculate the $\beta$ required to achieve a positive rate in the secret key model.

\begin{figure}
\includegraphics[width=\columnwidth]{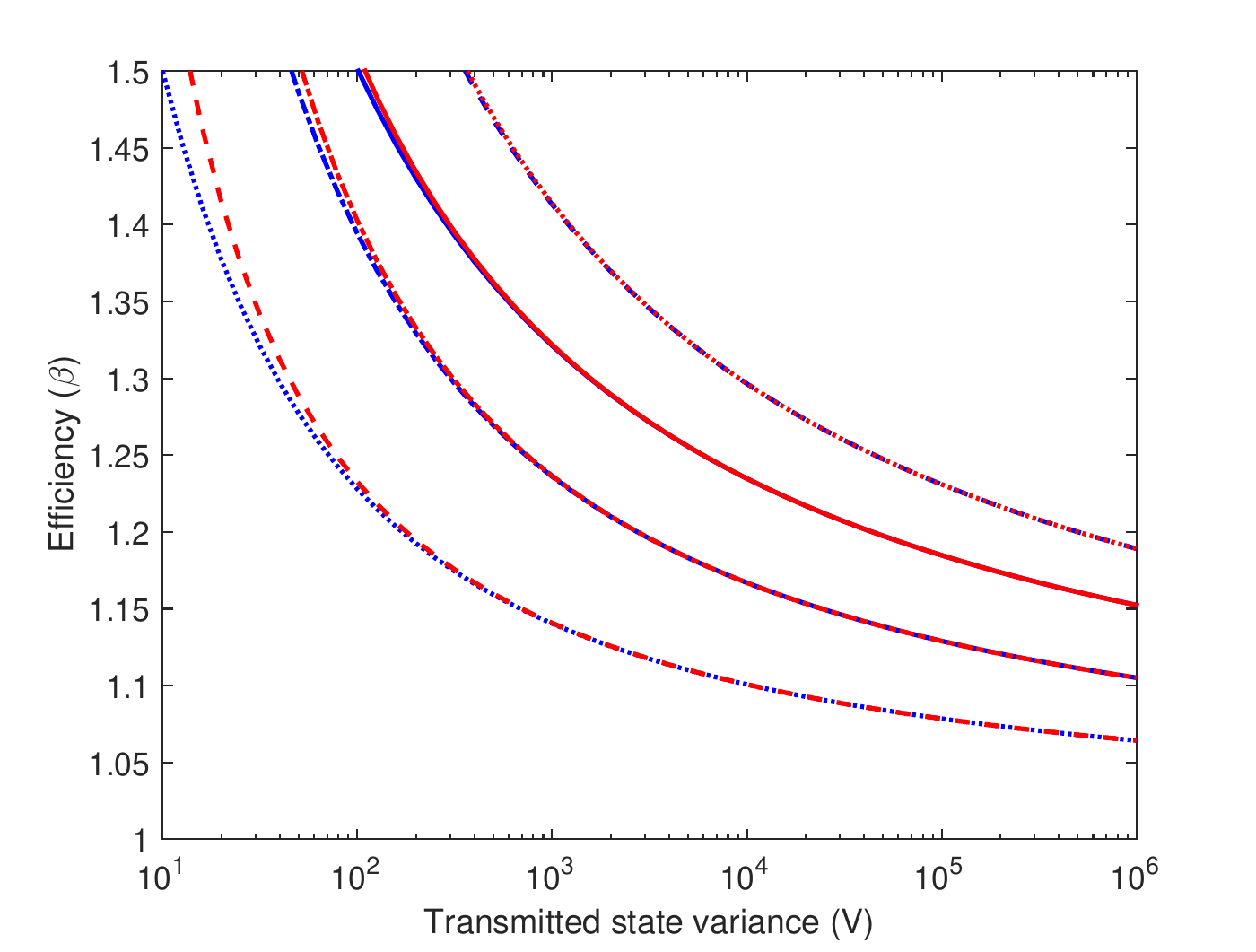}
\caption{
The $\beta$ required to achieve a positive
secret key rate for a unity-gain classical teleporter
quantum channel for various Gaussian CVQKD protocol configurations:
Direct reconciliation (blue) and reverse reconciliation (red);
Squeezed states and homodyne detection (solid line);
coherent states and homodyne detection (dotted line);
coherent states and heterodyne detection (dot-dashed line);
squeezed states and heterodyne detection (dashed line).
$\beta$ is plotted versus Alice's transmitted state
variance $(V)$.
}
\label{fig:beta-theory}
\end{figure}

Fig.~\ref{fig:beta-theory} shows the value of $\beta$ required to achieve a positive secret key for the following CVQKD protocols:
coherent or squeezed state sources;
homodyne or heterodyne detection;
and direct and reverse reconciliation protocols
(For details see Ref~\cite{Garcia_Patron_Sanchez_thesis2007}).
In Fig.~\ref{fig:beta-theory}, we have assumed collective attacks,
asymptotic key lengths, and ideal detection efficiency.
An example of the equations describing the channel capacity ($I_{\rm AB}$)
and Eve's maximum accessible information ($I_{\rm E}$) are detailed in Appendix~1.
Fig.~\ref{fig:beta-theory} plots the required $\beta$ verses Alice's transmitted state variance ($V=V_{\rm A}+1$), which is the only free parameter and is normalized to the quantum noise limit.

Fig.~\ref{fig:beta-theory} shows that $\beta$s
only slightly greater than unity are required to achieve a
positive key rate for a number of Gaussian CVQKD protocols
operating over unity-gain classical teleporter quantum channel.
In the previous section we demonstrated that the
secret key rate model Eq.~\eqref{eq:Leverrier_key_rate_2}
can lead to greater than unity $\beta$.
This is in addition to recent advances in Gaussian to binary mapping
protocols have demonstrated high efficiencies at low-SNR ratios
(For example see Ref~\cite{Leverrier_2008}).
Finally, we note that greater $\beta_{\rm FEC}$
values could be obtained by operating poorer FEC codes,
poorer in the sense of a slower drop off in error rate
with increasing SNR above capacity, as these codes can also
have a slower increase in error rate
with decreasing SNR below capacity. \\

 \section{A discussion on the secret key rate model}

In the previous section we demonstrated that the secret key rate models
given by in Eq.~\eqref{eq:key_rate} and Eq.~\eqref{eq:key_rate_short}
give incorrect results when employing fixed-rate forward error
correction and operating over a range of high WERs.
It is logical to reason that the secret key rate models
may then be incorrect for all non-zero WERs.
And since applications of fixed-rate ME-LDPC codes are operated at quite high
WERs, it is logical to conclude that the secret
key model is incorrect when such codes are used in the the error
reconciliation procedure
\cite{Lodewyck2007_LDPC_CVQKD,Fossier_CVQKD_NJP2009,Leverrier_2009,Jouguet-2011-multiedge,Jouguet2012_experimental,Huang_CVQKD2016}.

In the following we discuss how the current secret key rate model may be incorrect. \\

\subsection{A CVQKD post-selection perspective} \label{sec:Beta-prime}


Here we make the observation that the act of choosing which codewords to keep or discard based on their decoding performance is equivalent to a form of post-selection. In a general CVQKD post-selection protocol, Alice and Bob discard a subset of their data in-order to gain
an information advantage over Eve. Likewise, in an error reconciliation
process, Alice and Bob ``post-select'' the transmitted words that
decoded to correct codewords and discard those that did not.

A secret key rate model has been proposed in the context of the CVQKD post-selection
protocol \cite{Silberhorn_2002}.
The secret key rate model for a general protocol with post-selection is
\cite{Leverrier_phd2009}
\begin{equation} \label{eq:Postselection_key_rate}
\Delta I_{\rm PS} = f \beta I_{\rm AB} - I_{\rm E}
\vspace{-0.5em}
\end{equation}
where $f$ is the fraction of post-selected data. This secret
key rate model is not known to be tight and can be treated
a pessimistic lower bound \cite{Leverrier_phd2009}.
In the forward error correction context, the selected data
is the set of codewords that have been decoded to a valid codeword
and so a failure rate of $p_{\rm fail}$ results in the
fraction $f = 1 - p_{\rm fail}$ of the transmitted codewords
being post selected, which gives a post selection key rate of
\begin{eqnarray} \label{eq:new_key_rate}
\Delta I^\prime = (1-p_{\rm fail}) \beta I_{\rm AB} - I_{\rm E}.
\end{eqnarray}
An interpretation of this model is that all of
Eve's information is retained and distilled into
remaining key bits after error correction in the
case of the finite failure probability of the
forward error correction system.

We consider the performance of ME-LDPC codes as
the reconciliation step in a CVQKD system as
described in \cite[Figure~5]{Jouguet-2011-multiedge} using
both the traditional secret key model Eq.~\eqref{eq:key_rate} and the
post-selection secret key model Eq.~\eqref{eq:new_key_rate}.

\begin{figure}[h!]
    \centering
    \includegraphics[width=\columnwidth]{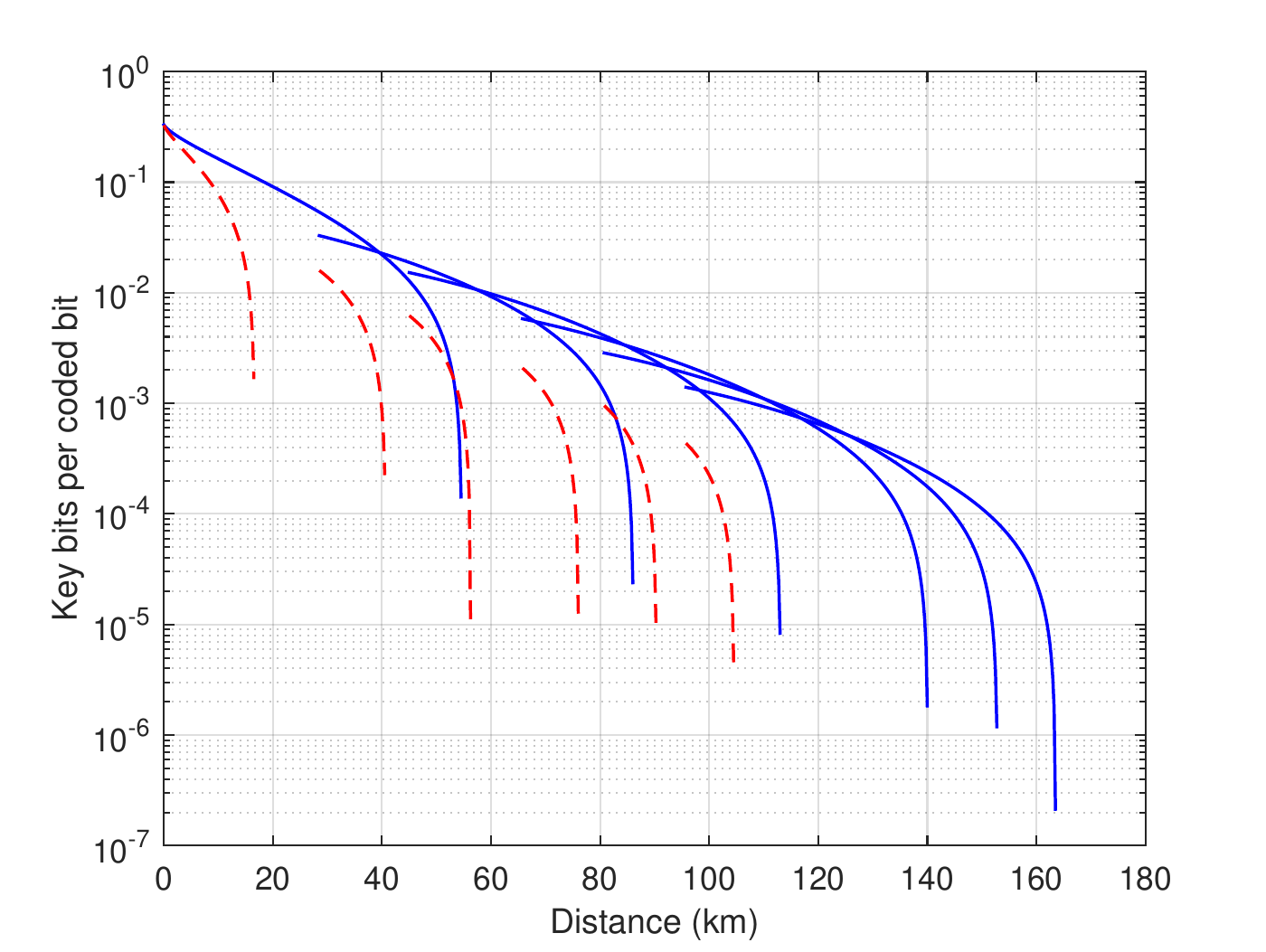}
    \caption{ Secret key models assuming collective attacks and employing
    ME-LDPC codes in the reconciliation procedure. The CVQKD protocol
    assumes reverse reconciliation, Gaussian modulated coherent
    states and homodyne detection. Same CVQKD parameters as Ref.~\cite{Jouguet-2011-multiedge}:
    $V_A\in\{1,100\}$
    modulation variance;
    $\varepsilon=0.01$ relative input channel excess noise;
    $\eta=0.6$ homodyne efficiency;
    $\nu_{\rm el}=0.01$ detector electronic noise;
    ${\rm WER}=1/3$.
    Solid blue: key rate model Eq.~\eqref{eq:key_rate},
    dashed red: key rate model Eq.~\eqref{eq:new_key_rate}.
    From right to left the ME-LDPC code parameters ($R,s,{\rm WER},\beta$),
    are \cite{Jouguet-2011-multiedge} $(0.005,0.00725,0.33,0.959)$,
    $(0.01,0.0145,0.33,0.966)$, $(0.02,0.029,0.33,0.969)$,
    $(0.05,0.075,0.33,0.958)$,
    $(0.1,0.0.161,0.33,0.931)$ and $(0.5,1.097,0.33,0.936)$. There is no dashed
    red curve shown for the $(0.005,0.00725,0.33,0.959)$ code as the key
    rate equation Eq.~\eqref{eq:new_key_rate} returns zero key bits for
    this code.
    \label{fig:QKD_new_key_rate}}
\end{figure}

Fig.~\ref{fig:QKD_new_key_rate} shows the
secret key models (Eq.~\eqref{eq:key_rate} and
Eq.~\eqref{eq:new_key_rate}) assuming collective attacks and
employing reverse reconciliation, Gaussian modulated
coherent states and homodyne detection. Both models
utilize six ME-LDPC codes in the reconciliation procedure
with performance data points ($R,s,{\rm WER},\beta$) as reported
in \cite{Jouguet-2011-multiedge} and choose the
value of signal variance, $1<V_{\rm A}<100$
corresponding to the given SNR, $s$. In short,
we have simply applied the same six data points
from \cite{Jouguet-2011-multiedge} to both
Eq.~\eqref{eq:key_rate} and Eq.~\eqref{eq:new_key_rate} using
the same CVQKD parameters in Ref.~\cite{Jouguet-2011-multiedge}.

Fig.~\ref{fig:QKD_new_key_rate} shows the impact of a high-WER
on both the secret key rate and operational range of the secret key
model Eq.~\eqref{eq:new_key_rate} with a high WER impacts both
the secret key rate and operational range of the protocol.
This suggests that it may be better to operate the FEC codes at a
much lower WER to maximize performance.

\begin{figure}[h!]
    \centering
    \includegraphics[width=\columnwidth]{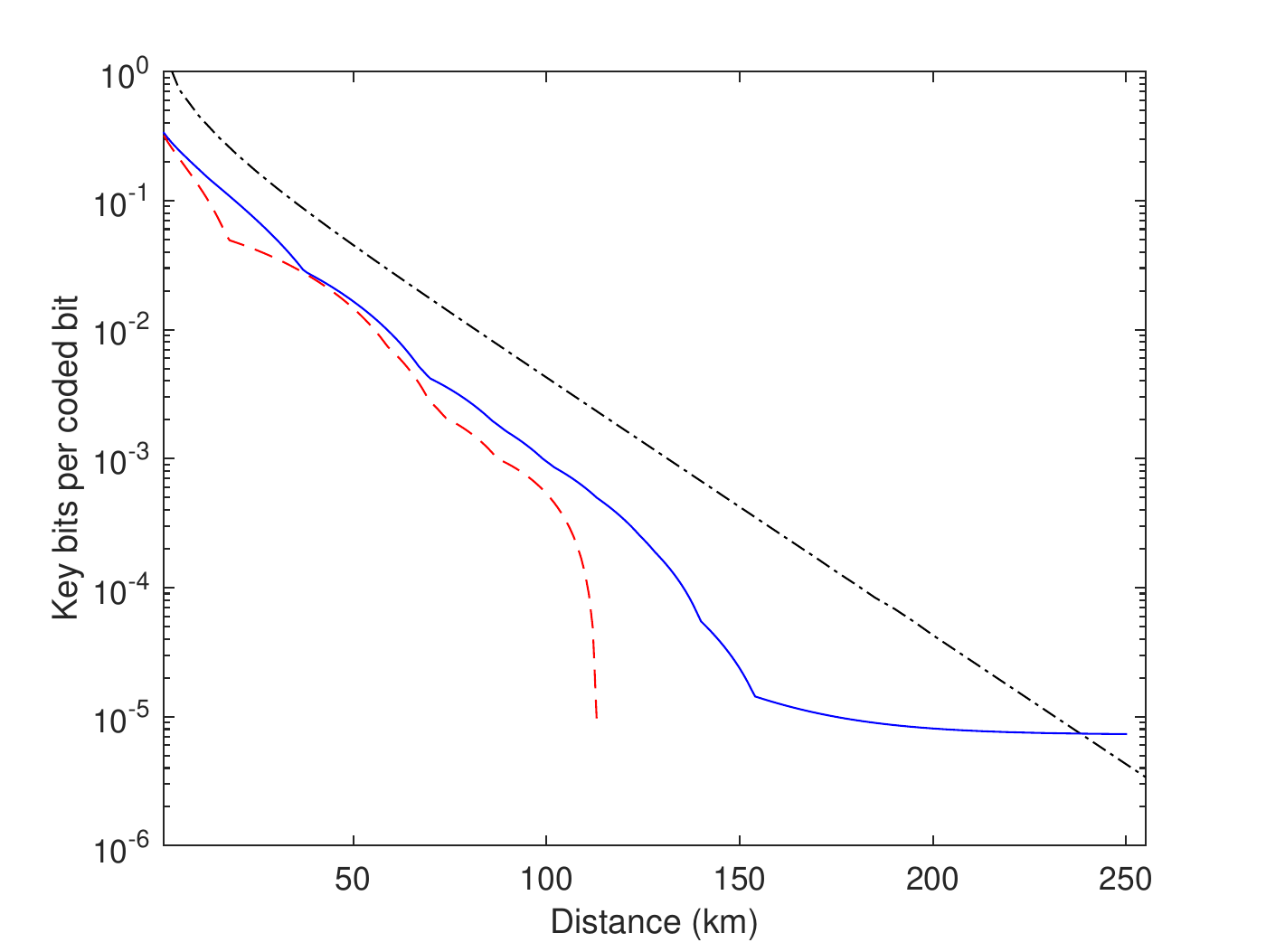}
    \caption{
    Secret key models assuming collective attacks and employing
    ME-LDPC codes in the reconciliation procedure. Same CVQKD parameters as
    Fig.~\ref{fig:QKD_new_key_rate}. The key rate models have been
    optimized over all ($R,s,$WER$,\beta$) points
    for the six ME-LDPC codes. The ME-LDPC codes with length $10^{5}$ bits
    and rates 0.5, 0.1, 0.05, 0.02, 0.01, and 0.005 are constructed randomly
    with degree distributions as given in \cite{Jouguet-2011-multiedge}.
    The dash-dot black curve gives the key rate for a theoretically optimal
    code $(\beta = 1,$ WER$ = 0)$. The solid blue curve gives the key
    rate calculated traditionally via Eq.~\eqref{eq:key_rate}
    while the dashed red curve gives the key rate calculated via Eq.~\eqref{eq:new_key_rate}.
    See Appendix~2 for more information on the data used to
    generate these curves. \label{fig:new_key_rate2}}
\end{figure}

To examine this further, Fig.~\ref{fig:new_key_rate2} shows the
effect on the key rate calculation for a set of ME-LDPC codes
when we jointly optimized over SNR, WER and code rate.
The CVQKD system is the same as described above for
Fig.~\ref{fig:QKD_new_key_rate}. We consider ME-LDPC codes
with length $10^{5}$ and rates 0.5, 0.1, 0.05, 0.02, and 0.005
with degree distributions as in \cite{Jouguet-2011-multiedge}.
For simplicity, we have assumed zero loss in efficiency due to
Gaussian to binary mapping, have ignored finite length effects
in all cases and have not placed any limitation on $V_{\rm A}$.
To obtain the values for SNR and WER for each code we simulated
their performance on the BI-AWGN channel over a range of
SNRs and WERs (see Appendix 2 for more detail on the optimization of
the parameters).

Fig.~\ref{fig:new_key_rate2} emphasizes the problem with the
secret key model Eq.~\eqref{eq:key_rate}. For high transmission losses,
the optimized key rate corresponds to a high-WER with $\beta$ above $1$. This corresponds to a better key rate than a theoretically optimal code $(\beta = 1,$ WER$ = 0)$. In contrast, Fig.~\ref{fig:new_key_rate2} shows the secret key model Eq.~\eqref{eq:new_key_rate}. The secret key rate for this model is optimized at significantly lower WERs with codes operating at lower $\beta$ values. We emphasize that the secret key model Eq.~\eqref{eq:new_key_rate} is a conservative model of the secret key rate.

\subsection{Operating at low word error rates} \label{sec:IT}

Increasing the WER allows us to increase the code rate $R$ above $I_{AB}$ thereby giving a positive
\[\beta I_{AB} - I_E = R - I_E,
\]
term even when $I_{AB} <  I_E$. The multiplicative correction term $(1-$WER) simply scales down the key rate leaving it positive.

\begin{figure}[h!]
    \centering
    \includegraphics[width=\columnwidth]{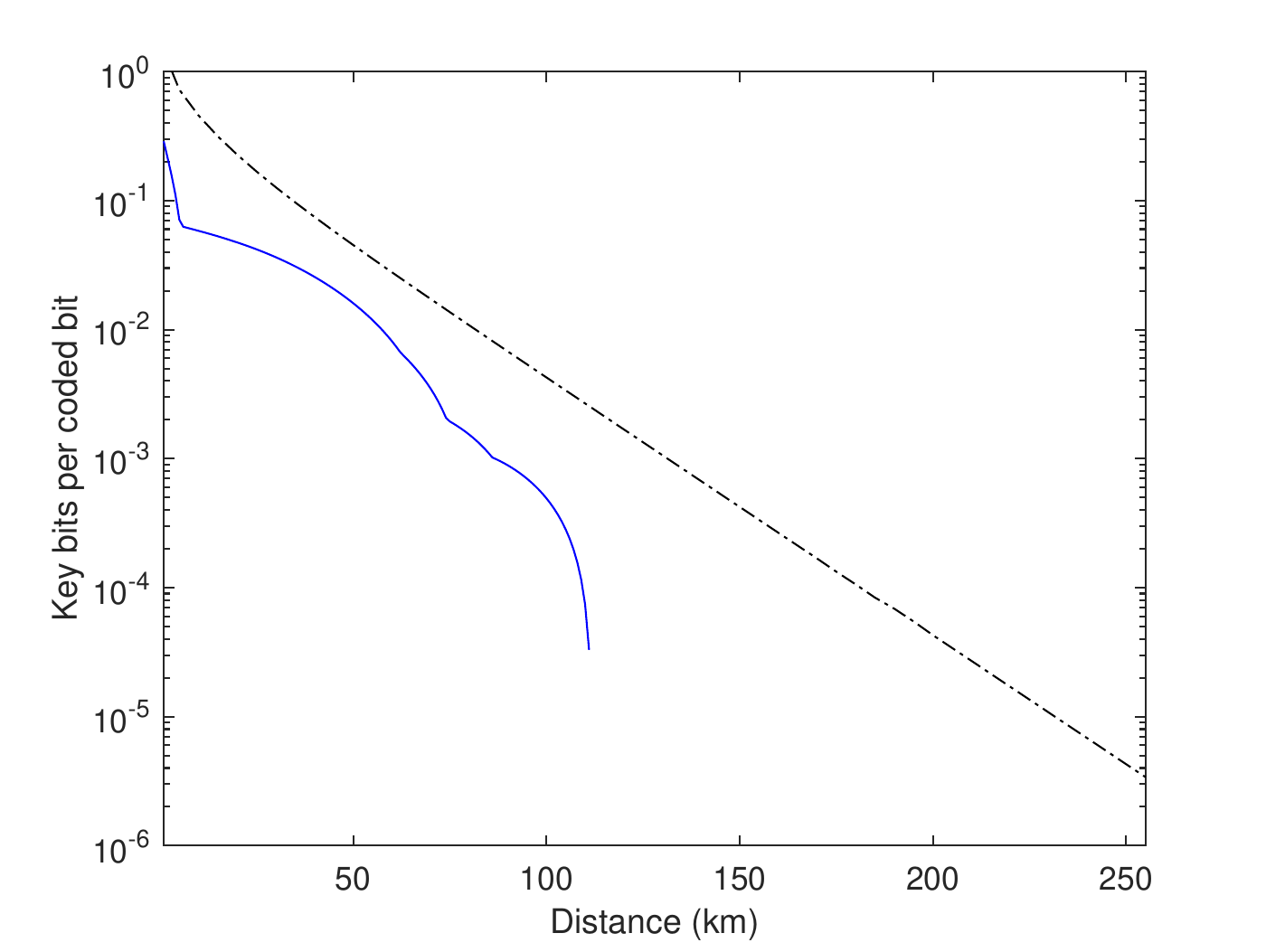}
    \caption{
    Secret key models assuming collective attacks and employing
    ME-LDPC codes in the reconciliation procedure. Same CVQKD parameters as
    Fig.~\ref{fig:QKD_new_key_rate}. The key rate models have been
    optimized over all ($R,s,$WER$,\beta$) points such that $\rm{WER}<0.05$
    for the six ME-LDPC codes. The ME-LDPC codes with length 100,000 bits
    and rates 0.5, 0.1, 0.05, 0.02, 0.01, and 0.005 are constructed randomly
    with degree distributions as given in \cite{Jouguet-2011-multiedge}.
    The dash-dot black curve gives the key rate for a theoretically optimal
    code $(\beta = 1,$ WER$ = 0)$. The solid red curve gives the key
    rate calculated traditionally via Eq.~\eqref{eq:key_rate}.
\label{fig:key_rate_reducedWER}}
\end{figure}

As we showed in Fig.~\ref{fig:ME-code-rate-1-50-results-v2c}, the error rate permitted has a significant effect on the code rate that can be achieved if it is allowed to be large. Of interest would be the operation of the forward error correction codes at error rates low enough so that the difference between the two cases is negligible. For ME-LDPC codes this means operating each code at a lower SNR or equivalently operating a much lower rate code at the same SNR. As an example, Fig.~\ref{fig:key_rate_reducedWER} shows the key rate we can obtain by using the same codes as the previous examples but limiting their operation to SNRs where the WER is below 0.05. Despite still allowing a quite high WER, a significant key rate loss is observed. This indicates that the high WER allowed previously played an important role in reporting good key rates.

While it may be possible, through good code design, to improve the code WER performance of ME-LDPC codes to some extent, for fixed rate codes there will always be a trade-off between operating at a SNR that returns a good efficiency and operating at an SNR that returns a low word error rate.
Nevertheless, there are alternative forward error correction strategies that do not use fixed-rate forward error correction codes. Instead we can employ so-called rateless Raptor codes that adjust the code rate in real time so as to always decode to a valid codeword, returning codes with both low word error rates and high efficiencies.


Raptor codes are graph based codes formed from the
concatenation of a high rate LDPC code with a Luby
Transform (LT) code. LT codes \cite{Luby} have very
simple encoding and decoding processes and can approach
the capacity of binary erasure channels with an unknown
erasure rate. The encoding and decoding of Raptor codes
are linear in terms of the message length; thus practical
for applications with large data transmission. Raptor
codes were studied for AWGN
channels in \cite{RaptorBSC}, where a systematic
framework was proposed to find the optimal degree distribution
across a range of SNRs. The
design of very low rate Raptor codes was studied in
\cite{Mahyar_TCOM2016_Raptor}. Using Raptor codes, a
potentially limitless number of coded symbols can be
generated, allowing the receiver to decode the message
once a sufficient number of parity bits have been
received; thus always decoding to a valid codeword.

\begin{figure}[h!]
    \centering
    \includegraphics[width=\columnwidth]{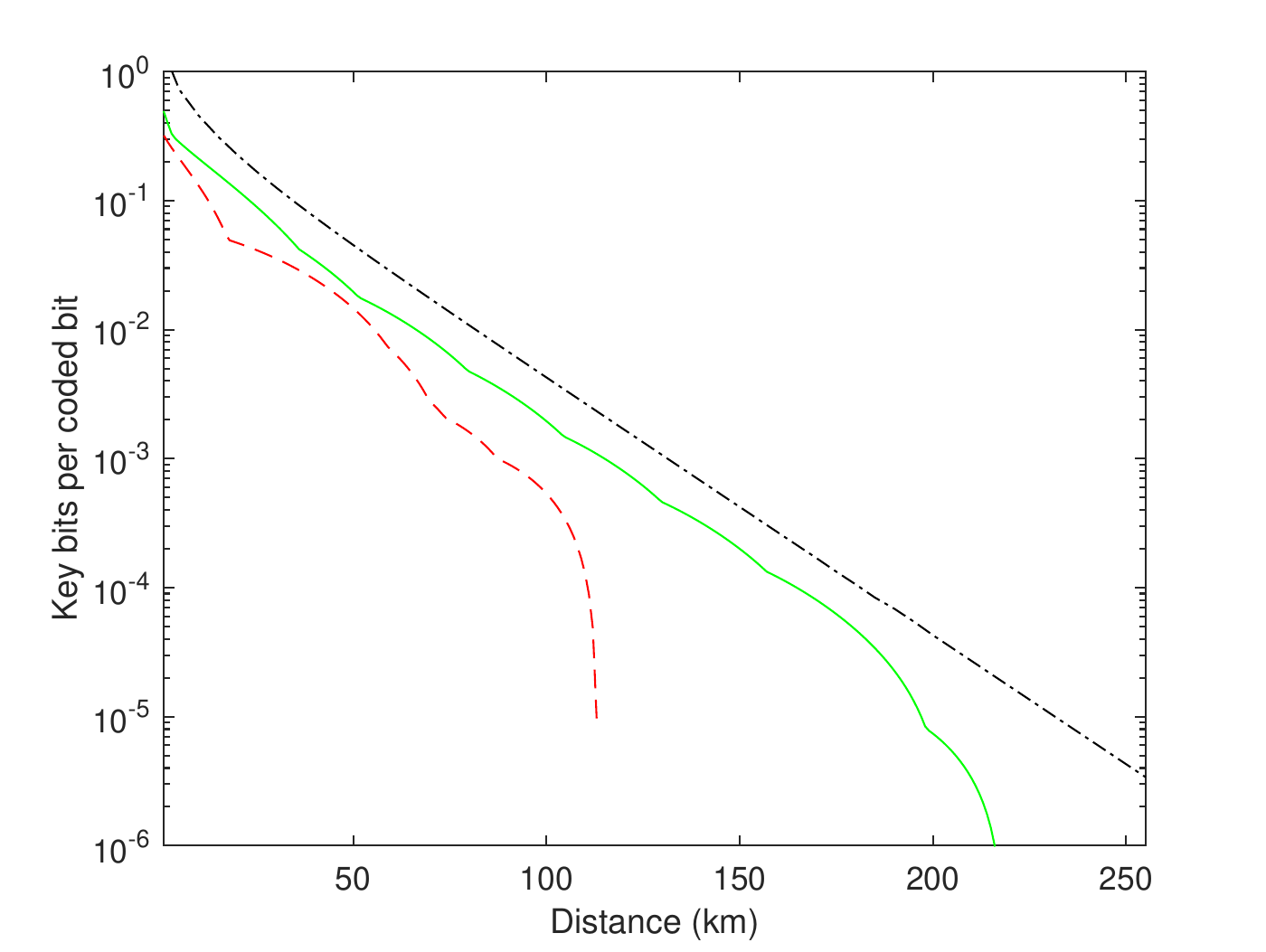}
    \caption{
    Secret key models assuming collective attacks and employing
    for Raptor and ME-LDPC codes in the reconciliation procedure.
    Same CVQKD parameters as Fig.~\ref{fig:QKD_new_key_rate}.
    For both ME-LDPC and Raptor codes, the secret key rate model
    has been optimized over the parameters ($R,s,$WER$,\beta$).
    The ME-LDPC codes with codeword length $n = 10^{5}$ bits
    and rates 0.5, 0.1, 0.05, 0.02, 0.01, and 0.005 are constructed
    randomly with degree distributions as given in
    \cite{Jouguet-2011-multiedge}. The raptor codes with message
    length $k=10^{4}$ are constructed randomly with degree
    distributions as given in Appendix 3. The dash-dot black curve
    gives the key rate for a theoretically optimal code $(\beta = 1,$ WER$ = 0)$.
    Solid green curve: the key rate model for Raptor codes via Eq.~\eqref{eq:new_key_rate}.
    Dashed red curve: key rate model for ME-LDPC codes via Eq.~\eqref{eq:new_key_rate}.
    \label{fig:key_rate_raptor}}
\end{figure}

It has been shown \cite{Mahyar_ICC2016_Raptor} that low-rate
Raptor codes can achieve higher efficiencies in comparison
with the fixed rate ME-LDPC codes in the entire SNR range
and do so at very low  WERs. When applied
to the reconciliation step of CVQKD, Raptor codes can
significantly improve the key rate \cite{Mahyar_ICC2016_Raptor}.
For example, Figure~\ref{fig:key_rate_raptor} shows the
key rate of the same CV-QKD system as considered in
Fig.~\ref{fig:QKD_new_key_rate} applying the Raptor
codes from \cite{Mahyar_ICC2016_Raptor,RaptorBSC}
(See Appendix~3 for more detail). For the Raptor codes the two models Eq.~\eqref{eq:new_key_rate} and Eq.~\eqref{eq:key_rate} return the same key-rate.

\subsection{A new model}

A full solution to this problem is non-trivial because the model involves finite-size effects (ie imperfect error correction) and asymptotic quantities. Such a solution would require a model of the number of bits revealed during the reconciliation procedure based on the particular code used, and how this quantity behaves in the asymptotic limit.

Protocols with a complete security proof in the finite-size regimes exist, for instance \cite{Furrer_PhysRevLett2012}. Such proofs for reconciliation using forward error correction will require a model of the number of bits leaked during the forward error correction reconciliation procedure as a function of the WER, and any uncertainty in it, as well as the codeword length and measurement errors.

\section{Conclusion} \label{sec:conclusion}

The maximum operation range of continuous variable
quantum key distribution systems has shown to be
improved by employing high-efficiency forward error
correction codes such as ME-LDPC codes.
In the current literature, CVQKD protocols with
fixed-rate forward error correction codes typically use a modified secret key
model Eq.~\eqref{eq:key_rate} that includes the WER term.

In this paper, we have demonstrated that this secret key model is
incorrect. We demonstrated this in two steps. Firstly, we showed that previously used
ME-LDPC codes for a range of high word error rates,
exhibited efficiencies greater than unity.
Secondly, we showed that assuming that code efficiencies could be
greater than unity, then using the secret key model Eq.~\eqref{eq:key_rate},
it was possible to achieve a positive secret key over an
entanglement breaking channel, which is equivalent to an intercept and
resend attack  - an impossible scenario.
We concluded that if the secret key model is incorrect
for a range of high WERs, it is also possibly incorrect
for any non-zero WERs.

We subsequently discussed the secret key model from the perspective of CVQKD post-selection protocols. In a CVQKD post-selection protocol, Alice and Bob discard
a subset of their data in-order to gain an information advantage over
Eve. Similarly, in an error reconciliation process, Alice and Bob
``post-select'' the code words that decoded to valid
codewords and discard codewords that they were unable to decode. This secret key rate model
is not known to be tight but can be treated as a
lower bound \cite{Leverrier_phd2009}. We showed that using the post selection key rate model
reduces the previously reported operational range of such
CVQKD systems employing fixed length forward error correction codes.
We also showed that using the current secret key rate model, but restricting the codes to operate at lower WERs, can also reduce the previously reported operational range of CVQKD systems employing fixed length forward error correction codes.
However, we did show that it is possible to employ an alternative forward error correction coding solution in the form of Raptor codes, which provide high efficiencies while operating at very low WERs.

A full solution to this problem would require a model of the number of bits revealed during the reconciliation procedure based on the particular code used, and how this quantity behaves in the asymptotic limit. Until such a model is known we would suggest that key rates obtained using the current model while operating at high WERs may not be accurate.

\subsection{Acknowledgments} \label{sec:acknowledgments}

Financial support was provided to A.M.L, T.S \& S.J.J. through the Australian Research Council Linkage Projects (Project ID LP130100783). S.J.J and L.O. are supported by the Australian Research Council via Future Fellowships (FT110100195 and FT140100219 respectively).

\section{Appendix 1} \label{sec:Appendix1}

In this section we detail Alice and Bob's channel capacity
as well as Eve's information for the Gaussian protocol with
reverse reconciliation and employing coherent states and
homodyne detection. Here we assume collective attacks and
asymptotic key lengths. The free parameters in these equations
are the following: $V$ variance of the transmitted state;
$T$ quantum channel with transmission;
$\varepsilon$ relative input channel excess noise;
$\eta$ homodyne efficiency;
$\nu_{\rm el}$ detector electronic noise.
Alice and Bob's channel capacity $(I_{\rm AB})$ and Eve's information $(I_{\rm E})$are
described by the following compact set of equations \cite{Fossier_CVQKDopt_2009}.

\begin{eqnarray}
I_{\rm AB}&=&C=\frac{1}{2}\log_2\bigg(\frac{V+\chi_{\rm tot}}{\chi_{\rm tot}+1}\bigg)\\
V&=&V_{\rm A}+1\\
\chi_{\rm tot}&=&\chi_{\rm line}+\chi_{\rm hom}/T \\
\chi_{\rm line}&=&1/T-1+\varepsilon \\
\chi_{\rm hom}&=&(1-\eta+\nu_{\rm el})/\eta \\
I_{\rm E}&=&G[(\lambda_1-1)/2]+G[(\lambda_2-1)/2]\\
& & -G[(\lambda_3-1)/2]-G[(\lambda_4-1)/2]\\
G[x]&=&(x+1)\log_2(x+1)-x\log_2(x)\\
\lambda_1&=&\sqrt{{\frac{1}{2}\bigg(A+\sqrt{A^2-4B}\bigg)}}\\
\lambda_2&=&\sqrt{{\frac{1}{2}\bigg(A-\sqrt{A^2-4B}\bigg)}}\\
\lambda_3&=&\sqrt{{\frac{1}{2}\bigg(C+\sqrt{C^2-4D}\bigg)}}\\
\lambda_4&=&\sqrt{{\frac{1}{2}\bigg(C-\sqrt{C^2-4D}\bigg)}}\\
A&=&V^2(1-2T)+2T+T^2(V+\chi_{\rm line})^2\\
B&=&T^2(V~\chi_{\rm line}+1)^2\\
C&=&\frac{A\chi_{\rm hom}+V\sqrt{B}+T(V+\chi_{\rm line})}{T(V+\chi_{\rm line})}\\
D&=&\sqrt{B}\frac{V+\sqrt{B}\chi_{\rm hom}}{T(V+\chi_{\rm line})}\\
\end{eqnarray}

The complete set of Gaussian protocol configurations, includes:
squeezed states and homodyne detection;
coherent states and heterodyne detection and
squeezed states and heterodyne detection. These protocols are
described in detail elsewhere \cite{Garcia_Patron_Sanchez_thesis2007}
in the case of ideal homodyne detector efficiency and
ideal forward error correction codes.

\section{Appendix 2} \label{sec:Appendix2}

\begin{figure}[h!]
    \centering
    \includegraphics[width=\columnwidth]{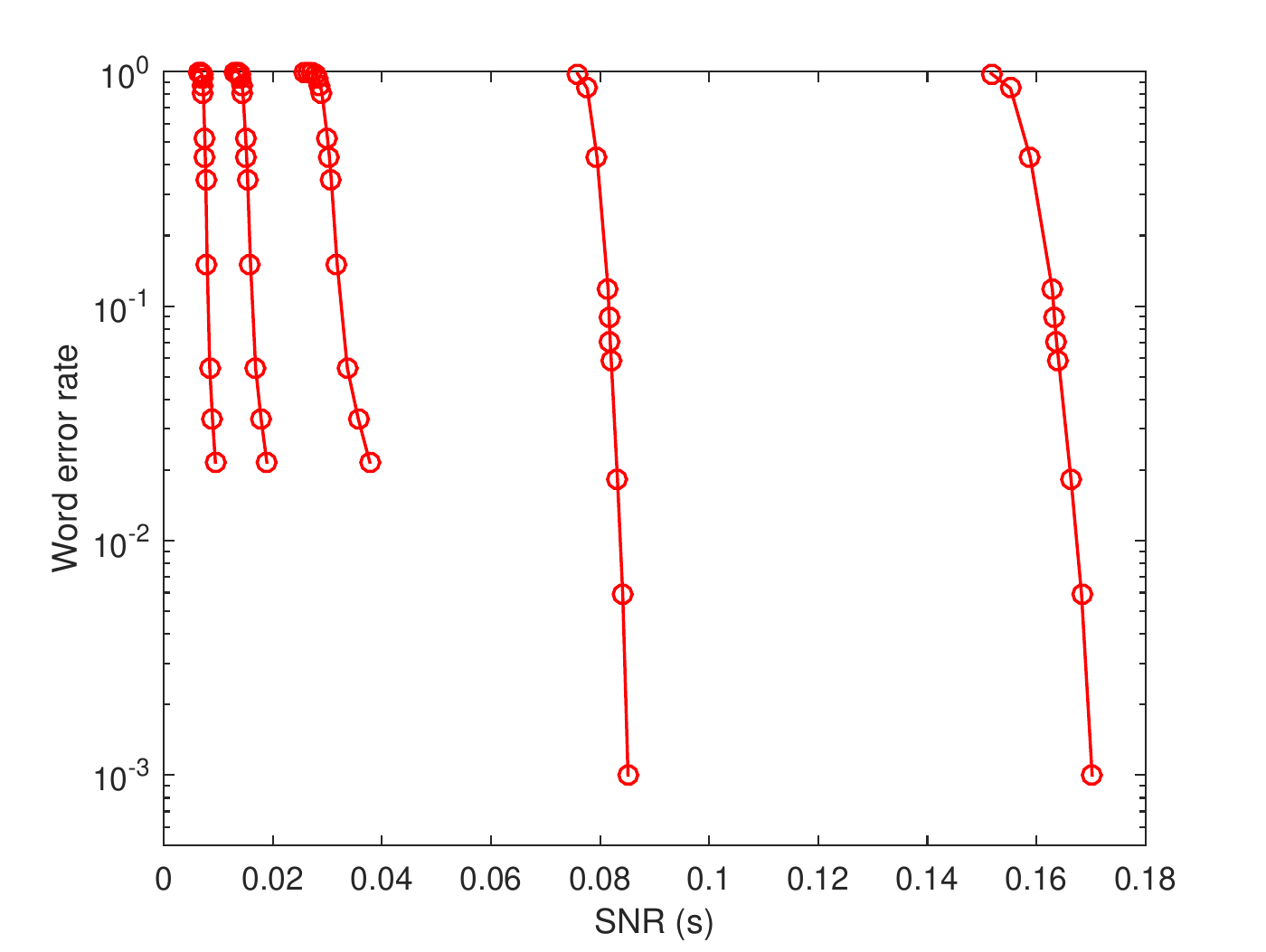}
\caption{ WER versus SNR for the considered
ME-LDPC codes From left to right are the
ME-LDPC codes with rates
0.005, 0.01, 0.02, 0.05, 0.1.
\label{fig:QKD_ME_error_curves}}
\end{figure}

In this section we provide details for optimizing the
secret key rate models shown in Fig.~\ref{fig:new_key_rate2}.
We consider ME-LDPC codes with length 100,000 and
rates 0.5, 0.1, 0.05, 0.02, and 0.005 with degree
distributions as in \cite{Jouguet-2011-multiedge}.
The parity-check matrices were constructed randomly
subject to the degree distribution constraints.
Fig.~\ref{fig:QKD_ME_error_curves} shows the performance of
the ME-LDPC codes with rates 0.005, 0.01, 0.02, 0.05, 0.1 as
the channel SNR is varied.

\begin{figure}[h!]
    \centering
    \includegraphics[width=\columnwidth]{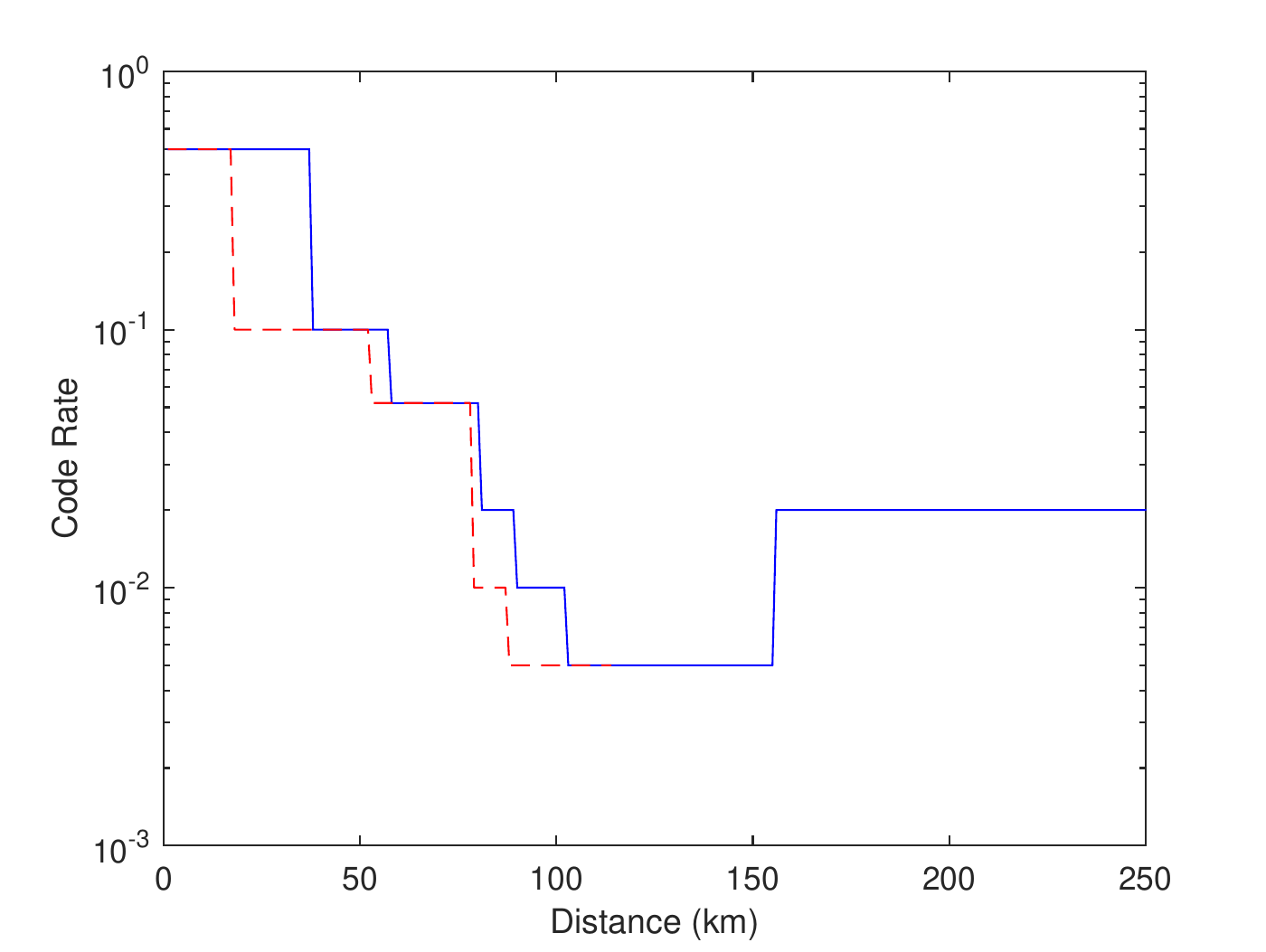}
\caption{Optimized code rate versus distance for
key rate equations Eq.~\eqref{eq:key_rate}
(solid blue) compared to key rate equations
Eq.~\eqref{eq:new_key_rate} (dashed red curve)
\label{fig:new_key_rate_R}}
\end{figure}

\begin{figure}[h!]
    \centering
    \includegraphics[width=\columnwidth]{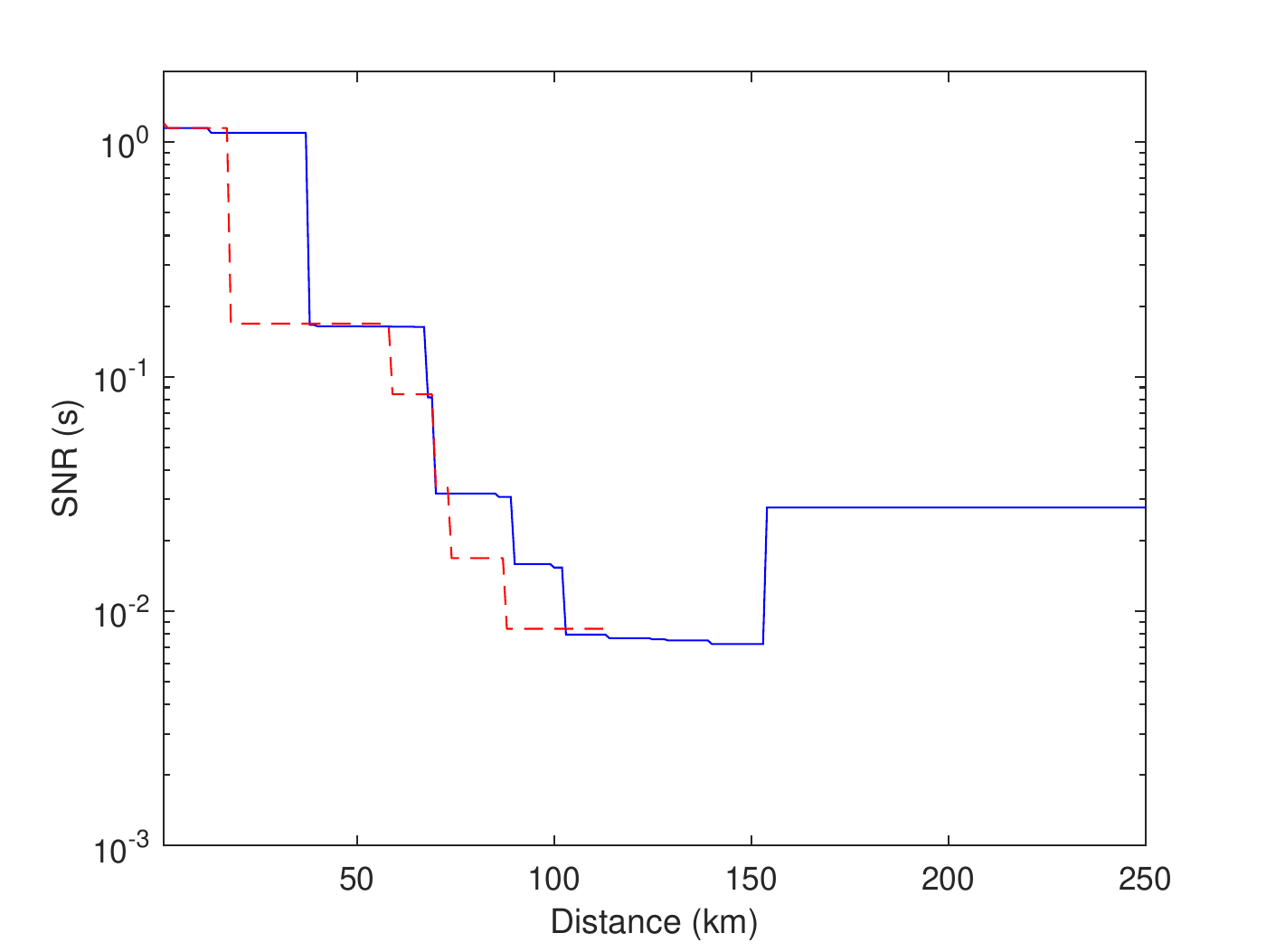}
\caption{Optimized SNR  versus distance
for key rate equations Eq.~\eqref{eq:key_rate}
(solid blue) compared to key rate equations
Eq.~\eqref{eq:new_key_rate} (dashed red curve)
\label{fig:new_key_rate_SNR}}
\end{figure}

\begin{figure}[h!]
    \centering
    \includegraphics[width=\columnwidth]{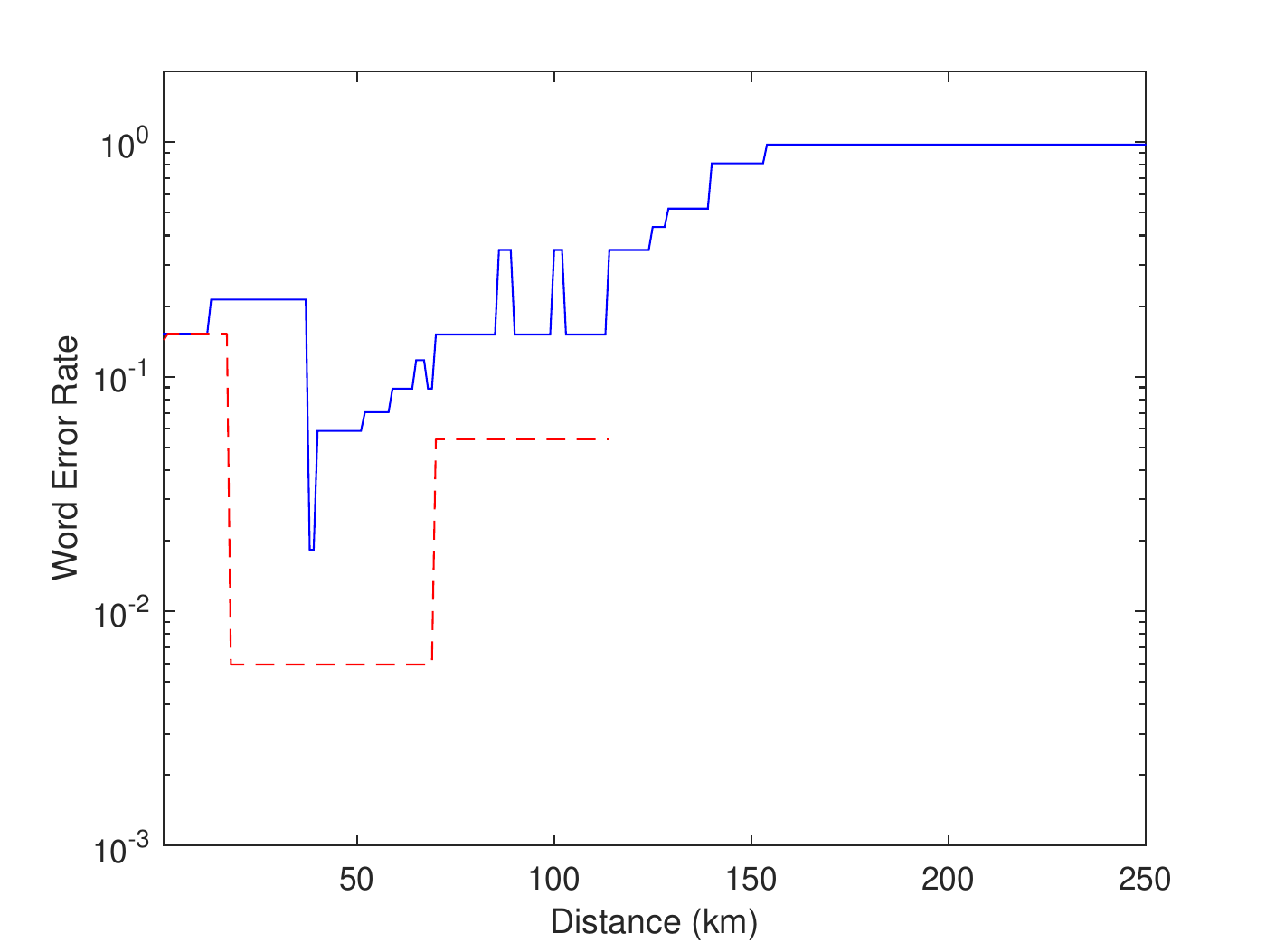}
    \caption{Optimized WER versus distance
    for key rate equations Eq.~\eqref{eq:key_rate}
    (solid blue) compared to key rate equations
    Eq.~\eqref{eq:new_key_rate}  (dashed red curve)
    \label{fig:new_key_rate_WER}}
\end{figure}

\begin{figure}[h!]
    \centering
    \includegraphics[width=\columnwidth]{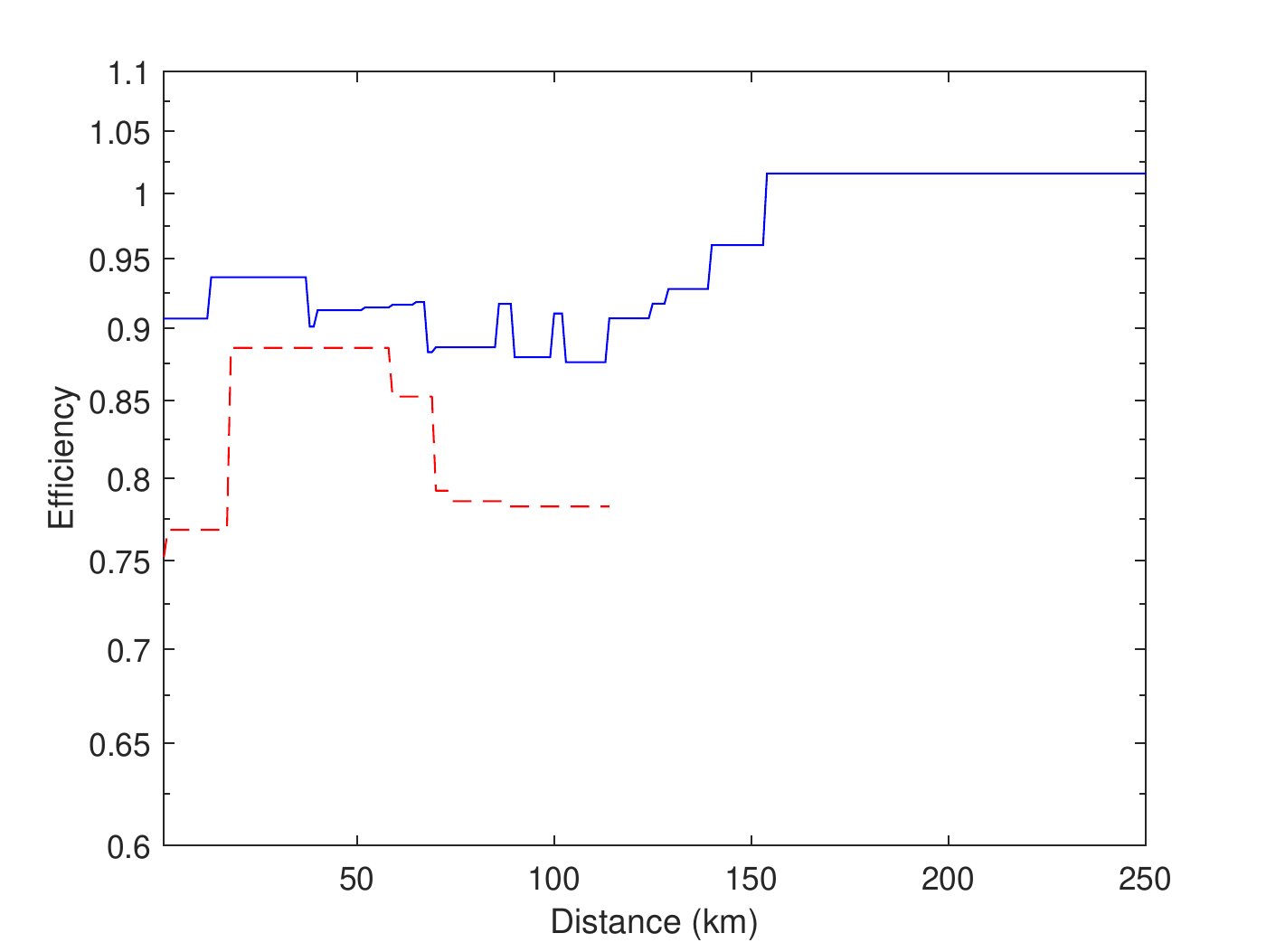}
 \caption{Optimized $\beta$ versus distance
 for key rate equations Eq.~\eqref{eq:key_rate}
 (solid blue) compared to key rate equations
 Eq.~\eqref{eq:new_key_rate} (dashed red curve)
 \label{fig:new_key_rate_beta}}
\end{figure}

We then jointly optimized over SNR, WER and rate
to maximize the secret key rate for each transmission
distance. The final secret key rate is the maximum
key rate over all codes. In Figs.~\ref{fig:new_key_rate_R}
to \ref{fig:new_key_rate_beta}, the solid red curves
give the parameters which optimize the key rate
calculated via Eq.~\eqref{eq:key_rate} while
the dashed red curves give the parameters which
optimize the key rate calculated via Eq.~\eqref{eq:new_key_rate}.

\section{Appendix 3} \label{sec:Appendix3}

We consider two Raptor codes in this paper. The first code,
from \cite{Mahyar_TCOM2016_Raptor}, has degree distribution
\begin{eqnarray}  \label{eqn:Raptor1}
\nonumber \Omega(x) &=& 0.0035x + 0.3538x^2 + 0.2337x^3 + 0.0737x^4 \\
\nonumber    &+& 0.0755x^5 + 0.0262x^6 + 0.0608x^7 + 0.0493x^{11} \\
 \nonumber   &+& 0.0255x^{12}+0.0002x^{21} + 0.0454x^{23}  \\
             &+& 0.0072x^{57}+0.0180x^{58}+0.0272x^{300}
\end{eqnarray}
The second Raptor code, from \cite{RaptorBSC}, has degree distribution
\begin{eqnarray}  \label{eqn:Raptor2}
\nonumber \Omega(x) &=&  0.0146 x^{1} + 0.3766x^{2} +  0.0677x^{3} +    0.2946x^{4} \\
\nonumber    &+& 0.1291x^{9} +   0.0060x^{12} +   0.0341x^{24} +   0.0228x^{29} \\
   &+&  0.0073x^{43} +   0.0472x^{200}
\end{eqnarray}

\begin{figure}[h!]
    \centering
    \includegraphics[width=\columnwidth]{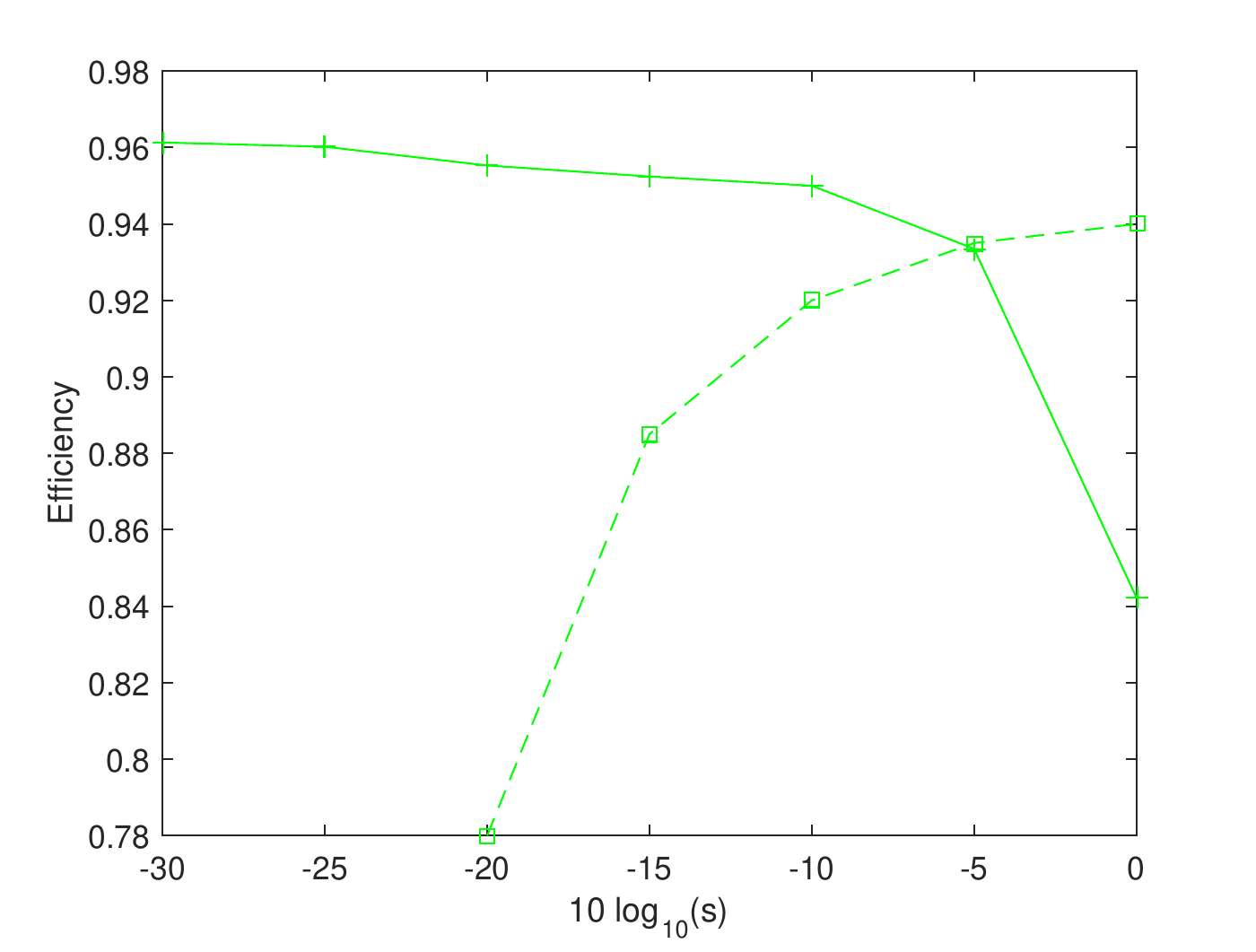}
 \caption{$\beta$ versus SNR for the two raptor codes.
 The solid curve gives $\beta$ values obtained for the
 code from \cite{Mahyar_TCOM2016_Raptor} with
 degree distribution \eqref{eqn:Raptor1} and message
 length $k = 100,000$. The dashed curve gives the
 efficiency of the code from from \cite{RaptorBSC}
 with degree distribution \eqref{eqn:Raptor2} and
 message length $k = 38,000$. \label{fig:Raptor_efficiency}}
\end{figure}

See \cite{Mahyar_TCOM2016_Raptor,RaptorBSC} for details
on encoding and decoding algorithms for these codes.
Figure~\ref{fig:Raptor_efficiency} shows the simulated
efficiency for these codes on an AWGN channel as the SNR is varied.

\end{document}